\documentclass[aps,preprint]{revtex4}%
\usepackage{amsfonts}
\usepackage{amsmath}
\usepackage{amssymb}
\usepackage{graphicx}%
\begin{document}
\title{\textbf{Noncommutative information is revealed from Hawking radiation as
tunneling }}
\author{Baocheng Zhang$^{1,2}$}
\email{zhangbc@wipm.ac.cn}
\author{Qing-yu Cai$^{1}$}
\author{Ming-sheng Zhan$^{1,2}$}
\author{Li You$^{3}$}
\affiliation{$^{1}$State Key Laboratory of Magnetic Resonances and Atomic and Molecular
Physics, Wuhan Institute of Physics and Mathematics, The Chinese Academy of
Sciences, Wuhan 430071, PRC}
\affiliation{$^{2}$Center for Cold Atom Physics, Chinese Academy of Sciences, Wuhan 430071,
PRC }
\affiliation{$^{3}$Department of Physics, Tsinghua University, Beijing 100084, PRC}

\begin{abstract}
We revisit the tunneling process from a Schwarzschild black hole in the
noncommutative spacetime and obtain the non-thermal tunneling probability. In
such non-thermal spectrum, the correlations are discovered, which can carry
the information about the noncommutativity. Thus this enlightens a way to find
the noncommutative information in the Hawking radiation. The entropy is also
shown to be conserved in the whole radiation process, which implies that the
unitarity is held even for the Hawking radiation from noncommutative black holes.

PACS classification codes: 04.70.Dy, 03.67.-a, 02.40.Gh

Keywords: Tunneling; Noncommutativity; Correlation; Entropy Conservation

\end{abstract}
\maketitle

\section{\textbf{Introduction}}

Hawking's semiclassical analysis \cite{swh75,swh76} of the black hole
radiation suggests that the information collapsed into the black hole will
lose for ever since the thermal radiation can not carry any information. This
means that the unitarity as required by quantum mechanics is violated.
However, it is found that the background geometry is considered fixed and the
energy conservation is not enforced during the radiation process in Hawking
method. Recently, Parikh and Wilczek suggested \cite{pw00} a method based on
energy conservation by calculating the particle flux in Painlev\'{e}
coordinates from the tunneling picture. Their result recovered the Hawking's
original result in leading order and gave the consistent temperature
expression and the entropy relation. The method had also been discussed
generally in different situations \cite{rzg05,amv05,jwc06,ma06,gs10} and
showed the formula was self consistent even when checked by using
thermodynamic relation \cite{sk08,zcz08,bms08,zcz090,jkp09}. Another important
aspect is to give the non-thermal spectrum, as shown in Ref. \cite{zcz09} that
there exist information-carrying correlation in the radiation spectrum and the
entropy is conserved in the sequential tunneling process. Along this line the
extention has been made in Ref. \cite{zcz091,cs09,sv10,iy10}.

It is noted that the noncommutativity had been introduced into the
investigation of Hawking radiation as tunneling \cite{bms08}. The
noncommutativity \cite{anss07,mkp07,ms08,bmm09} can provide the minimal length
scale upon the generalized uncertainty principle since the Heisenberg
uncertainty principle may not be satisfied when quantum gravitational effects
become important. Moreover, the noncommutativity also provides a totally
different black hole and the noncommutative black hole thermodynamics is also
investigated in the Parikh-Wilczek tunneling picture \cite{bms08}.
Particularly, the noncommutativity leads to the only remnant result of black
hole radiation since it can remove the so-called Hawking paradox where the
temperature diverges as the radius of a standard black hole shrinks to zero.
This is advantageous over the standard Schwarzschild black hole whose final
radiation will lead to divergence of the temperature and over the quantum
corrected black hole whose final radiation is dependent on the quantum
corrected parameter which may be different in string theory and loop quantum
gravity theory [5]. In the paper, we will show there exist correlations among
the radiated particles in the situation of noncommutative black holes and
these correlations could carry the information about noncommutativity hidden
in spacetime. We also check the entropy conservation in the radiation process,
which is consistent with the unitarity of quantum mechanics.

The organization of the paper is as follows. In the second section we revisit
the tunneling through the noncommutative black hole and discuss its
thermodynamics. The third section is devoted to investigation of correlation
and entropy conservation for the radiation of noncommutative black hole.
Finally, we summarize our results in the fourth section.

In this paper we take the unit convention $k=\hbar=c=G=1$.

\section{Tunneling in noncommutative space and thermodynamics}

In the section we will recalculate the particles' tunneling probability from
the Schwarzschild black hole in noncommutative space, along the line presented
in Ref. \cite{bms08}. In order to include the noncommutative effect in
gravity, we can change the mass of gravitating object. The usual definition of
mass density in commutative space is expressed in terms of Dirac delta
function, but in noncommutative space the form breaks down due to
position-position uncertainty relation. It is shown that noncommutativity
eliminates point-like structures in favor of smeared objects in flat
spacetime. The effect of smearing is implemented by redefining the mass
density by a Gaussian distribution of minimal width $\sqrt{\theta}$ instead of
the Dirac delta function. Here $\theta$ is the noncommutative parameter which
is considered to be a small (Planck length) positive number and comes from the
noncommutator of $\left[  x^{\mu},x^{\upsilon}\right]  =i\theta^{\mu\nu}$ with
$\theta^{\mu\nu}=\theta diag\left[  \epsilon_{1},\cdots,\epsilon_{D/2}\right]
$. It is noted the constancy of $\theta$ is related to a consistent treatment
of Lorentz invariance and unitarity. For the purpose of noncommutativity, the
mass density is chosen as%

\begin{equation}
\rho_{\theta}\left(  r\right)  =\frac{M}{\left(  4\pi\theta\right)  ^{\frac
{3}{2}}}\exp\left(  -\frac{r^{2}}{4\theta}\right)  ,\label{nmd}%
\end{equation}
which plays the role of a matter source and the mass is smeared around the
region $\sqrt{\theta}$ instead of locating at a point. To find a solution of
Einstein equation $G_{\mu\nu}=8\pi T_{\mu\nu}$ with the noncommutative mass
density of the type (\ref{nmd}), the energy-momentum tensor is identified as
$T_{\nu}^{\mu}=diag[-\rho_{\theta},q_{r},q_{t},q_{t}]$ which provides a
self-gravitating droplet of anisotropic fluid with the radial pressure
$q_{r}=-\rho_{\theta}$ and tangential pressure $q_{t}=-\rho_{\theta}-\frac
{1}{2}r\partial_{r}\rho_{\theta}=\left(  \frac{r^{2}}{4\theta}-1\right)
\frac{M}{\left(  4\pi\theta\right)  ^{\frac{3}{2}}}\exp\left(  -\frac{r^{2}%
}{4\theta}\right)  $. It is seen easily that the pressure is anisotropic, but
at the large values of $r$ all the components of the energy-momentum tensor
tend to zero very quickly and so the pressure is again isotropic and the
Schwarzschild vacuum solution is well applicable.

Solving the Einstein equation in the noncommutative space leads to the solution,%

\begin{equation}
ds^{2}=-\left(  1-\frac{4M}{r\sqrt{\pi}}\gamma\left(  \frac{3}{2},\frac{r^{2}%
}{4\theta}\right)  \right)  dt^{2}+\left(  1-\frac{4M}{r\sqrt{\pi}}%
\gamma\left(  \frac{3}{2},\frac{r^{2}}{4\theta}\right)  \right)  ^{-1}%
dr^{2}+r^{2}d\Omega^{2} \label{nsc}%
\end{equation}
where the lower incomplete gamma function is defined by%

\begin{equation}
\gamma\left(  \frac{3}{2},\frac{r^{2}}{4\theta}\right)  =\int_{0}^{\frac
{r^{2}}{4\theta}}t^{\frac{1}{2}}e^{-t}dt \label{igf}%
\end{equation}
Note that when $r$ goes to infinity, $\gamma$ approaches to $\sqrt{\pi}/2$.
Comparing the the noncommutative coordinates (\ref{nsc}) with the commutative
one \cite{pw00}, we note that their difference lies in the mass term, that is
to say, substituting the mass term of Schwarzschild spacetime by $m_{\theta
}=\int_{0}^{r}4\pi r^{^{\prime}2}\rho_{\theta}\left(  r^{^{\prime}}\right)
dr^{^{\prime}}=\frac{2M}{\sqrt{\pi}}\gamma\left(  \frac{3}{2},\frac{r^{2}%
}{4\theta}\right)  $. It is noted that in the $\theta\longrightarrow0$ limit,
the incomplete $\gamma$ function becomes the usual gamma function and
$m_{\theta}\left(  r\right)  \longrightarrow M$ that is the commutative limit
of the noncommutative mass $m_{\theta}\left(  r\right)  $. From the condition
of $g_{tt}(r_{h})=0$, the event horizon can be found as $r_{h}=\frac{4M}%
{\sqrt{\pi}}\gamma\left(  \frac{3}{2},\frac{r_{h}^{2}}{4\theta}\right)
\equiv\frac{4M}{\sqrt{\pi}}\gamma_{h}$. Keeping up to the leading order
$\frac{1}{\sqrt{\theta}}e^{-M^{2}/\theta}$, we find $r_{h}\simeq2M(1-\frac
{2M}{\sqrt{\pi\theta}}e^{-M^{2}/\theta})$.

In what follows, in order to describe cross-horizon phenomena of tunneling
particles, we have to change the coordinates (\ref{nsc}) to quasi-Painlev\'{e}
coordinates, which are regular and not singular at the horizon. Doing the time
Painlev\'{e} coordinate transformation, we obtain the new coordinates as%

\begin{equation}
ds^{2}=-\left(  1-\frac{4M}{r\sqrt{\pi}}\gamma\right)  dt^{2}+2\left(
1-\frac{4M}{r\sqrt{\pi}}\gamma\right)  \sqrt{\frac{\frac{4M}{r\sqrt{\pi}%
}\gamma}{\left(  1-\frac{4M}{r\sqrt{\pi}}\gamma\right)  ^{2}}}dtdr+dr^{2}%
+r^{2}d\Omega^{2} \label{nspc}%
\end{equation}
where $\gamma\equiv\gamma\left(  \frac{3}{2},\frac{r^{2}}{4\theta}\right)  $
and the spacetime described by (\ref{nspc}) is still stationary. The radial
null geodesics are obtained by setting $ds^{2}=d\Omega^{2}=0$ in (\ref{nspc}),%

\begin{equation}
\overset{.}{r}=\pm1-\sqrt{\frac{4M}{r\sqrt{\pi}}\gamma} \label{nsg}%
\end{equation}
where the upper (lower) sign can be identified with the outgoing (incoming)
radial motion, under the implicit assumption that time t increases towards the future.

Let us consider a positive energy shell to cross the horizon in the outward
direction from $r_{i}$ to $r_{f}$. Along the method given by Parikh and
Wilczek, the imaginary part of the action for that shell is given by%

\begin{equation}
\operatorname{Im}I=\operatorname{Im}\int_{r_{i}}^{r_{f}}p_{r}%
dr=\operatorname{Im}\int_{r_{i}}^{r_{f}}\int_{0}^{p_{r}}dp_{r}^{^{\prime}%
}dr=\operatorname{Im}\int_{0}^{H}\int_{r_{i}}^{r_{f}}\frac{dr}{\overset{.}{r}%
}dH^{^{\prime}} \label{nsa}%
\end{equation}

Moreover, if the self-gravitation is included, we have to make the replacement
$M\longrightarrow M-E$ in Eqs. (\ref{nspc}) and (\ref{nsg}), where $E$ is the
outgoing particle's energy. Thus the expression (\ref{nsa}) is modified as%

\begin{equation}
\operatorname{Im}I=\operatorname{Im}\int_{M}^{M-E}\int_{r_{i}}^{r_{f}}%
\frac{drd\left(  M-E\right)  }{\overset{.}{r}}=-\operatorname{Im}\int_{0}%
^{E}\int_{r_{i}}^{r_{f}}\frac{dr}{\overset{.}{r}}dE^{^{\prime}}%
\end{equation}
where we have changed the integration variable from $H^{^{\prime}}$ to
$E^{^{\prime}}$. Then inserting the Eq. (\ref{nsg}) into the expression of the
imaginary part of the action, we have%

\begin{equation}
\operatorname{Im}I=-\operatorname{Im}\int_{0}^{E}\int_{r_{i}}^{r_{f}}\frac
{dr}{1-\sqrt{\frac{4M}{r\sqrt{\pi}}\gamma}}dE^{^{\prime}}%
\end{equation}
The $r$-integration is done by deforming the contour. A detailed calculation
gives the tunneling rate,%

\begin{equation}
\Gamma_{N}\left(  E\right)  \sim e^{-2\operatorname{Im}I}=\exp\left(  -8\pi
E\left(  M-\frac{E}{2}\right)  +16\sqrt{\frac{\pi}{\theta}}M^{3}e^{\frac
{M^{2}}{\theta}}-16\sqrt{\frac{\pi}{\theta}}\left(  M-E\right)  ^{3}%
e^{\frac{\left(  M-E\right)  ^{2}}{\theta}}\right)  \label{nst}%
\end{equation}
where the result is obtained up to the leading order $\frac{1}{\sqrt{\theta}%
}e^{-M^{2}/\theta}$.

In what follows we will check the thermodynamics for tunneling from
noncommutative black hole. Generally, we can obtain the temperature from the
tunneling probability by comparing it with the thermal Boltzmann relation.
Here it is difficult to expanding the expression (\ref{nst}) to obtain the
leading term. However we could also consider the first law of thermodynamics
to check whether the radiation temperature given by thermodynamic relation
$\frac{1}{T}=\frac{dS}{dM}$ is equal to $\frac{\kappa}{2\pi}$ where $\kappa$
is the surface gravity of the black hole. Since the metric (\ref{nsc}) is
static, the surface gravity can be calculated as $\kappa=\frac{1}{2}%
\frac{dg_{tt}}{dr}|_{r=r_{h}}=\frac{1}{2}\left[  \frac{1}{r_{h}}-\frac
{r_{h}^{2}}{4\theta^{3/2}}\frac{e^{-\frac{r_{h}^{2}}{4\theta}}}{\gamma_{h}%
}\right]  $. Thus one can get the temperature,%

\begin{equation}
T_{N}=\frac{\kappa}{2\pi}=\frac{1}{4\pi r_{h}}\left[  1-\frac{r_{h}^{3}%
}{4\theta^{3/2}}\frac{e^{-\frac{r_{h}^{2}}{4\theta}}}{\gamma_{h}}\right]  .
\label{nstt}%
\end{equation}
When the noncommutative parameter $\theta\longrightarrow0$, the temperature
decays into the standard form, $T_{H}=\frac{1}{8\pi M}$. Note that for
standard form of black hole temperature, in the limit $M\longrightarrow0$, the
temperature will be infinite. But the consideration of noncommutative black
hole radiation will avoid the divergence problem. From the expression
(\ref{nstt}), we obtain that the temperature will fall down to zero at some
definite value, i.e. $r_{h}=r_{0}$. As a result, the noncommutativity
restricts evaporation process to a remnant. In the region of $r_{h}<r_{0}$,
there is no black hole since the temperature can not be defined
\cite{mkp07,bmm09}.

On the other hand, we can obtain the entropy of the noncommutative black hole
by deforming the Bekenstein-Hawking relation,%

\[
S_{N}=\pi r_{h}^{2}=\frac{16M^{2}}{\pi}\gamma_{h}^{2}\simeq4\pi M^{2}%
-16\sqrt{\frac{\pi}{\theta}}M^{3}e^{-\frac{M^{2}}{\theta}}%
\]
One can check easily that the temperature given by $\frac{1}{T}=\frac{dS}{dM}$
is equivalent to that given by (\ref{nstt}). More importantly, we can compare
the tunneling probability (\ref{nst}) with the Boltzmann factor $e^{-E/T}$ and
find that the tunneling probability (\ref{nst}) gives a non-thermal spectrum.
Moreover, a detailed calculation shows that $\Delta S_{N}=S_{Nf}%
-S_{Ni}=-2\operatorname{Im}I$ with $r_{i}=2M(1-\frac{2M}{\sqrt{\pi\theta}%
}e^{-M^{2}/\theta})$ and $r_{f}=2\left(  M-\omega\right)  (1-\frac{2\left(
M-\omega\right)  }{\sqrt{\pi\theta}}e^{-\left(  M-\omega\right)  ^{2}/\theta
})$ up to the leading order $\frac{1}{\sqrt{\theta}}e^{-M^{2}/\theta}$. So we have,%

\begin{equation}
\Gamma=e^{\Delta S_{N}}%
\end{equation}
which shows the tunneling probability is related to the change of
noncommutative black hole entropy.

\section{Correlation and entropy conservation in the radiation process}

In the previous section it has been shown that the tunneling probability
satisfied the relation $\Gamma=e^{\Delta S}$ and gave a non-thermal spectrum
for the noncommutative black hole. Such a spectrum is intriguing and could
give some suggestions for the black hole information loss paradox. In the
section, along the line outlined by us earlier \cite{zcz09}, we will show that
there exists correlation between the tunneling particles and the entropy is
conserved in the tunneling process.

Considering two emissions with energies $E_{1}$ and $E_{2}$ and using the
expression (\ref{nst}), we have%

\[
\Gamma_{N}\left(  E_{1}\right)  =\exp\left(  -8\pi E_{1}\left(  M-\frac{E_{1}%
}{2}\right)  +16\sqrt{\frac{\pi}{\theta}}M^{3}e^{\frac{M^{2}}{\theta}}%
-16\sqrt{\frac{\pi}{\theta}}\left(  M-E_{1}\right)  ^{3}e^{\frac{\left(
M-E_{1}\right)  ^{2}}{\theta}}\right)
\]

\begin{align*}
\Gamma_{N}\left(  E_{2}|E_{1}\right)   &  =\exp(-8\pi E_{2}\left(
M-E_{1}-\frac{E_{2}}{2}\right) \\
&  +16\sqrt{\frac{\pi}{\theta}}\left(  M-E_{1}\right)  ^{3}e^{\frac{\left(
M-E_{1}\right)  ^{2}}{\theta}}-16\sqrt{\frac{\pi}{\theta}}\left(
M-E_{1}-E_{2}\right)  ^{3}e^{\frac{\left(  M-E_{1}-E_{2}\right)  ^{2}}{\theta
})}%
\end{align*}
where the tunneling probability for the second radiation is a conditional
probability given the occurrence of tunneling of the particle with energy
$E_{1}$. According to the definition of joint probability in statistical
theory \cite{nc00}, we get%

\begin{align*}
\Gamma_{N}\left(  E_{1},E_{2}\right)   &  =\Gamma_{N}\left(  E_{1}\right)
\Gamma_{N}\left(  E_{2}|E_{1}\right) \\
&  =\exp(-8\pi\left(  E_{1}+E_{2}\right)  \left(  M-\frac{E_{1}+E_{2}}%
{2}\right) \\
&  +16\sqrt{\frac{\pi}{\theta}}M^{3}e^{\frac{M^{2}}{\theta}}-16\sqrt{\frac
{\pi}{\theta}}\left(  M-E_{1}-E_{2}\right)  ^{3}e^{\frac{\left(  M-E_{1}%
-E_{2}\right)  ^{2}}{\theta}})
\end{align*}
We can check that $\Gamma_{N}\left(  E_{1},E_{2}\right)  =$ $\Gamma_{N}$
$\left(  E_{1}+E_{2}\right)  $ which is the emission of particle with the
energy $E_{1}+E_{2}$.

In order to evaluate the statistical correlation which says that two events
are correlated if the probability of the two events arising simultaneously is
not equal to the product probabilities of each event occurring independently,
we have to integrate the $E_{1}$ variable in $\Gamma_{N}\left(  E_{1}%
,E_{2}\right)  $ to attain the independent probability $\Gamma_{N}\left(
E_{2}\right)  $,%

\begin{align*}
\Gamma_{N}\left(  E_{2}\right)   &  =\Lambda\int_{0}^{M-E_{2}}\Gamma_{N}%
(E_{1},E_{2})dE_{1}\\
&  =\exp\left(  -8\pi E_{2}\left(  M-\frac{E_{2}}{2}\right)  +16\sqrt
{\frac{\pi}{\theta}}M^{3}e^{\frac{M^{2}}{\theta}}-16\sqrt{\frac{\pi}{\theta}%
}\left(  M-E_{2}\right)  ^{3}e^{\frac{\left(  M-E_{2}\right)  ^{2}}{\theta}%
}\right)
\end{align*}
where $\Lambda$ is the normalized factor which is the function of the black
hole mass $M$, stemmed from the normalization of tunneling probability
$\Lambda\int\Gamma_{N}\left(  E\right)  dE=1$. Now we can calculate the
statistical correlation between the two radiations
\begin{equation}
C\left(  E_{1},E_{2},\theta\right)  =\ln\Gamma(E_{1}+E_{2})-\ln\left[
(\Gamma(E_{1})\ \Gamma(E_{2})\right]  \neq0.
\end{equation}
Thus the adoption of a non-commutative spacetime does not change our statement
that a non-thermal spectrum affirms the existence of correlation, as is
illustrated for a Schwarzschild black hole. We find that the information
associated with noncommutativity is factored out in the correlation even in
the early stage of Hawking radiation. Thus even though non-commutativity only
exists at the small scale, we can still test its effect through correlations
contained in the non-thermal spectrum of Hawking radiation. Especially, the
LHC experiment could produce the micro black hole \cite{cp10} and the
analogous black hole radiation experiment has been realized through the laser
\cite{bcc10} or BEC \cite{lib10}, so one can observe such radiations to check
whether there is the information about noncommutativity in them.

For tunneling of two particles with energies $E_{1}$ and $E_{2}$, we find the
entropy
\begin{align}
S_{N}(E_{1})=-\ln\Gamma_{N}(E_{1}) &  =8\pi E_{1}\left(  M-\frac{E_{1}}%
{2}\right)  -16\sqrt{\frac{\pi}{\theta}}M^{3}e^{-\frac{M^{2}}{\theta}}%
+16\sqrt{\frac{\pi}{\theta}}\left(  M-E_{1}\right)  ^{3}e^{-\frac{\left(
M-E_{1}\right)  ^{2}}{\theta}},\\
S_{N}(E_{2}|E_{1})=-\ln\Gamma_{N}(E_{2}|E_{1}) &  =8\pi E_{2}\left(
M-E_{1}-\frac{E_{2}}{2}\right)  \nonumber\\
&  -16\sqrt{\frac{\pi}{\theta}}\left(  M-E_{1}\right)  ^{3}e^{-\frac{\left(
M-E_{1}\right)  ^{2}}{\theta}}+16\sqrt{\frac{\pi}{\theta}}\left(
M-E_{1}-E_{2}\right)  ^{3}e^{-\frac{\left(  M-E_{1}-E_{2}\right)  ^{2}}%
{\theta}}.
\end{align}
It is seen easily that they satisfy the definition for conditional entropy
$S(E_{1},E_{2})=-\ln\Gamma(E_{1}+E_{2})=S(E_{1})+S(E_{2}|E_{1})$. A detailed
calculation confirms that the amount of correlation is exactly equal to the
mutual information described in Ref. \cite{zcz09}, and this shows that it is
the correlation that carries away the information. If we count the total
entropy carried away by the outgoing particles, we find
\begin{equation}
S_{N}(E_{1},E_{2},\cdots,E_{n})=\sum\limits_{i=1}^{n}S_{N}(E_{i}|E_{1}%
,E_{2},\cdots,E_{i-1}).
\end{equation}

Thus we show that entropy is conserved in Hawking radiation for a
Schwarzschild black hole in a noncommutative spacetime. Note that the
temperature will be zero before black hole vanishes, that is, the black hole
will evolve into a remnant which is in a high-entropy state with entropy
$S_{NC}=4\pi E_{NC}^{2}$. So finally the black hole entropy can be found as%

\[
S_{NB}=S_{N}(E_{1},E_{2},\cdots,E_{n})+S_{NC}%
\]

Especially if the reaction is included, the tunneling probability is%

\begin{align}
\Gamma_{NR}\left(  E\right)   &  \sim\left[  1-\frac{2E\left(  M-\frac{E}%
{2}\right)  }{M^{2}+\alpha}\right]  ^{-4\pi\alpha}\exp\left[  -8\pi E\left(
M-\frac{E}{2}\right)  \right]  \times\nonumber\\
&  \exp\left[  16\sqrt{\frac{\pi}{\theta}}M^{3}e^{-\frac{M^{2}}{\theta}%
}-16\sqrt{\frac{\pi}{\theta}}\left(  M-E\right)  ^{3}e^{-\frac{\left(
M-E\right)  ^{2}}{\theta}}+\mathrm{const.}\,(\mathrm{independent\text{ }%
of}\text{ }M)\right]  .\label{nbtp}%
\end{align}
Using the same method as above, we can also show that there exists
correlations among radiations and the entropy is conserved in the tunneling
radiation process. Our method for this case, however, does not solve the
remaining problem of whether the black hole will evaporate to exhaustion or
will halt at some value of a critical mass because of the reaction effect or
the use of a noncommutative spacetime. As we discussed before, when the
quantum reaction parameter $\alpha$ is negative, a black hole will leave
behind a remnant instead of radiating into exhaustion \cite{lx07}. If
noncommutative spacetime is introduced, however, the parameter $\alpha$ cannot
be negative, in order to avoid a divergent temperature at the end of
radiation. That is to say, when the non-commutative parameter $\theta\neq0$, a
reasonable value for $\alpha$ would be positive, and thus when the mass of the
black hole is reduced to a certain value, the temperature will drastically
decrease to zero to form an extreme black hole \cite{bms08}. Despite these
subtleties, we have shown in this section that the entropy is conserved in the
tunneling process and so the information could not be lost even for the
noncommutative Schwarzschild black hole by taking into account information
carried away by correlations in emitted particles.

\section{\textbf{Conclusion}}

Using the tunneling method and considering the noncommutative effect in
Schwarzschild spacetime, the modified tunneling probability is derived. Based
on this probability, we have shown that the adoption of a noncommutative
spacetime supports that there exist correlations in non-thermal spectrum and
the correlation can carry all information, even including the information
about the noncommutativity, which may be observed in the future LHC experiment
or simulating experiment in laboratory. The entropy conservation is also
investigated, which implies that the radiation process of black hole is
unitary in the background of noncommutative spacetime.

\section{Acknowledgement}

This work is supported by National Basic Research Program of China (NBRPC)
under Grant No. 2006CB921203 and NSFC under Grant No. 11074283.

\end{document}